\documentclass[aps,prl,twocolumn,a4paper,amsmath,amssymb,showpacs,%
superscriptaddress,floatfix,]{revtex4}

\usepackage{graphicx}
\usepackage{epstopdf}

\def\openone{\leavevmode\hbox{\small1\kern-3.3pt\normalsize1}}

\begin{document}

\title{Quantum quasiresonances in grazing incident angle 
atom-surface collisions} 
\date{\today}

\author{Antonia Ruiz}
\affiliation{Departamento de F\'{i}sica Fundamental y Experimental,
  Universidad de La Laguna, La Laguna 38204, Spain}
\author{Jos\'{e} P. Palao}
\affiliation{Departamento de F\'{i}sica Fundamental II,
  Universidad de La Laguna, La Laguna 38204, Spain}
\author{Eric J. Heller}
\affiliation{Departments of Chemistry and Physics,
  Harvard University, Cambridge, Massachusetts 02138, USA}

\begin{abstract}

The momentum transfer between the normal components to an index direction in the 
collision of an atom with a periodic surface is investigated. For fast atoms with 
grazing angle of incidence there is an interval of azimuthal angles around the index 
direction for which the energy transfer can be very efficient. 
This effect is reflected in quantum diffraction patterns with large 
non-specular peaks, associated with the parallel to the surface and normal 
to the index direction momentum component, and can be described in terms 
of quasiresonance.
Although the classical dynamics does not reproduce the precise quantum diffraction 
probabilities, indicating the quantum nature of this effect, classical and quantum 
computations show that the span of the quasiresonance region coincides in both 
dynamics and can be classically estimated from the phase-space analysis. 

\end{abstract}

\maketitle


Recent experiments in grazing angle of incidence collisions of fast atoms 
with a surface along a low index direction have shown quantum diffraction 
patterns even for a total translational energy as large as several keV
\cite{Rousseau07,Schuller07,Schuller08}.
This result is explained by the existence of two motion regimes, a fast
one along a low index direction, with momentum $p_x=\sqrt{2mE}\sin\theta\cos\varphi$, 
and a slow one in the plane perpendicular to the index direction, with momentum
components
$p_y=\sqrt{2mE}\sin\theta\sin\varphi$ and $p_z=\sqrt{2mE}\cos\theta$, where 
$E$ is the total incident energy and $m$ the atom mass.
The component $p_z$ is chosen normal to the surface. For incidence along an index 
direction, $\varphi_i=0$ and $p_{yi}=0$.
Within these conditions, the fast atoms ``feel'' the surface as structureless 
parallel furrows aligned along the index direction \cite{Rousseau07}, which 
constitutes an example of channelling on surfaces \cite{Schuller07,Gemmell74}.
As a consequence the two regimes of motion are decoupled, resulting in a constant 
momentum component $p_x$. Under grazing angle conditions the energy associated with 
the normal components is very small, and, as the change in $p_y$ is quantized, a 
quantum diffraction pattern is observed despite the relatively large total translational 
energy. 

However, the previous analysis does not explain to what extent the diffraction pattern
still survives for incidence not in a surface index direction ($p_{yi}\neq0$),  
nor does it predict how efficient is the energy transfer between the normal components 
responsible for the diffraction pattern. A convenient analysis of the issue can be made 
in terms of quasiresonance.
The concept of quasiresonance was first introduced to explain some very efficient and specific 
energy transfer between the vibrational and rotational degrees of freedom experimentally 
observed in state resolved collisions between vibrotationally excited diatom molecules and 
one atom \cite{Stewart88,Magill88}. 
An striking observation was a remarkable insensitivity of the quasiresonant effect to the
specific details of the interaction potential \cite{Stewart88,Magill89,Scott96}.
It was also shown that a fully classical treatment of the collision reproduces 
the experimental results remarkably well \cite{Magill88,Magill89,Hoving89,Forrey99,Stewart00}.
Recently, the quasiresonance analysis has been extended to new processes in different 
classical systems  \cite{Ruiz05,Ruiz06}, including grazing angle atom surface collisions 
\cite{Ruiz08}.

An important open question is whether the quantum effects in the dynamics share any 
common mechanism with the classical quasiresonances. This issue has been analyzed in 
ultracold atom-diatom collisions \cite{Forrey99}. The results were not conclusive 
as for the low energies considered the main inelastic quantum channels associated with 
quasiresonances remain closed.
Grazing angle atom-surface collisions in a close-to an index incident direction provides a 
new ground to study the similarities between both dynamics. 
%


%
Classically, quasiresonance can be understood in terms of the adiabatic invariance theory and the 
method of averaging \cite{Lichtenberg92}. Let us consider a two-dimensional integrable system 
described by the action variables $(J_1,J_2)$, perturbed by a interaction, depending on other 
additional degrees of freedom, which is switched on and off. The interaction induces a coupling 
between the two ``internal'' degrees of the perturbed system, and defines a process from an 
initial state $(J_{1i},J_{2i})$ to a final state $(J_{1f},J_{2f})$. 
Let us assume
(a) a slowly varying perturbative interaction; and 
(b) an initial internal state satisfying the approximate $M:N$ non-linear resonance condition 
$M\omega_{2i}-N\omega_{1i}\approx 0$, with $M$ and $N$ small integers, and $\omega_\alpha$ the two 
independent frequencies of the system.
The analysis of dynamics of the nearly resonant system in terms of the standard secular perturbation 
theory introduces a canonical transformation to new action variables, $I_1=-J_1/N,I_2=J_2+MJ_1/N$, 
defined in a rotating frame in which the angle variable $\phi_2$ associated with the action 
$I_2$ oscillates much more rapidly than the other variables during the entire transient process. 
The averaging of the transformed Hamiltonian over this rapidly oscillating angle leads to its
conjugated action as the first term of the series expansion for the adiabatic invariant of the 
system in the proximity of the nonlinear resonance. In terms of the original actions, the 
existence of this local adiabatic invariance implies that
\begin{equation}
\frac{\Delta J_2}{\Delta J_1}\,=\,-\frac{M}{N}\,,
\end{equation}
where $\Delta J=J_f-J_i$. This accurate correlation between the action changes is the central 
result of the quasiresonance analysis \cite{Ruiz05,Ruiz06}. 
The quasiresonance region is defined as the interval of initial values for which the conditions 
(a) and (b) are satisfied. 
%


In atom surface collisions, classical quasiresonance applies to an effective two-dimensional 
free atom, described by the components of the momentum parallel to the surface, which is perturbed by 
the transient atom-surface interaction. 
Here the analysis can be simplified considering a coordinate system with the x-axis fixed along
the index direction of interest, since in this coordinate system the actions $I_1$ and $I_2$ can be
directly identified with the components of the momentum $p_y$ and $p_x$.
The ``external'' translational action that controls the strength of the interaction is $p_z$.

Assumption (a) corresponds to a collision with large translational energy (perturbative regime) and a 
small $p_{z}$ (slowly variant interaction); both together implying $p_{z}<<p_{x}$. Meanwhile the 
assumption (b) corresponds to $p_{y}<<p_{x}$.  
These conditions define a scattering process with a grazing incident beam closely aligned 
to an index direction of the surface.
Since ${\dot x}>>{\dot y},\,{\dot z}$, the averaging over this rapidly oscillating 
variable $(\phi_2\equiv x)$ allows us to identify its conjugated action variable, 
the momentum along such direction $(I_{2}\equiv p_{x})$, as the first order adiabatic 
invariant characterizing quasiresonance. 
The dynamics in the perpendicular plane $(yz)$ is dictated by the Hamiltonian 
\begin{equation}\label{hav}
H_{av}=\frac{1}{2m}\left(p_y^2+p_z^2\right)+V_{av}(y,z)
\end{equation}
with the averaged potential
\begin{equation}\label{vav}
V_{av}(y,z)\,=\,\frac{1}{L}\int_0^{L}V(x,y,z)\,dx,
\end{equation}
being $L$ the surface periodicity along the x-axis.
In a quasiresonance region, the coupling induced by this averaged potential enables an 
energy transfer between the normal components of the momentum, $p_y$ and $p_z$.
The efficiency of such energy transfer is determined by the form of $V_{av}$
resulting from the original potential and the particular index direction implied. 
In general, the energy transfer is more efficient for incidence close to a low index 
direction.
Outside quasiresonance such energy transfer does not occurs, and incident beams 
``feel'' a plain surface and only suffer an effective specular reflection.

In a coordinate system with the x-axis fixed along the surface direction  $(1,0)$, the adiabatic 
invariance of the momentum along a symmetry direction $(M,N)$, associated with the approximated 
resonance condition $Mp^{\prime}_{y}-Np^{\prime}_{x}\simeq 0$, leads to the correlation
$\Delta p_y^\prime/\Delta p_x^\prime=-M/N$ \cite{Ruiz08}.
In this coordinate system the efficient energy transfer in the plane perpendicular to the 
direction $(M,N)$ is reflected by significant changes in both internal actions, 
$p^{\prime}_{x}$ and $p^{\prime}_{y}$. This is the usual result 
of the general quasiresonance analysis.


Quasiresonance is satisfied for grazing incident beams aligned over an interval of azimuthal
angles around the crystal index direction.
We define the width $W_{py}$ of the quasiresonance region as the interval of initial values of $p_{y}$ 
for which the approximated resonance condition $(b)$ is satisfied, or, alternatively, $W_\varphi$ as 
the corresponding interval of incident azimuthal angles $\varphi_i$.
Both quantities are related by $W_{\varphi}=2\sin^{-1}[W_{py}/(2\sqrt{2mE}\sin\theta_i)]$.
$W_{py}$ can be estimated from the analysis of the phase space associated with the averaged 
Hamiltonian at frozen values $z_{fr}$ of the normal $z$ coordinate \cite{Ruiz06}. 
We focus on the dynamics around the  resonant incident direction, $p_{yi}=0$, which can 
be properly described by the pendulum like Hamiltonian $H_{pend}=p_{y}^{2}/2m+V_{av}(y,z_{fr})$.   
The maximum excursion in action of the islands of the resonance zone in this integrable 
system provides the simple expression
\begin{equation}\label{wpy}
W_{py}\,\approx\,2\sqrt{2m\left[V_{av}(y_{u},z_{m})-V_{av}(y_{s},z_{m})\right]},
\end{equation}
being $z_{m}$ the normal coordinate that gives the maximum strength interaction at incident energy $E$,
$y_{u}$ the hyperbolic point associated with the innermost separatrix, and $y_{s}$ the most stable fixed
point of the potential $V_{av}(y,z_{m})$.
The expression (\ref{wpy}) is valid provided that the resonance zones associated with the different 
crystal index directions remain well isolated in phase space. In a non perturbative regime, the overlapping 
between neighbouring resonance zones in phase space makes the quasiresonance regions shrink 
\cite{Ruiz06}.


To illustrate quasiresonance we simulate the atom-surface collision considering a Lennard-Jones 
and Devonshire potential with parameters taken from Ref. \cite{Celli85}, describing the elastic scattering 
of He atoms from a perfectly periodic rigid LiF(001) surface.
The low normal velocities implicit under grazing angle conditions, and the robustness of the quasiresonance 
phenomena with respect to the details of the interaction potential justify the use of this simple model.


\begin{figure}[tb]\centering
\includegraphics[width=0.9\linewidth]{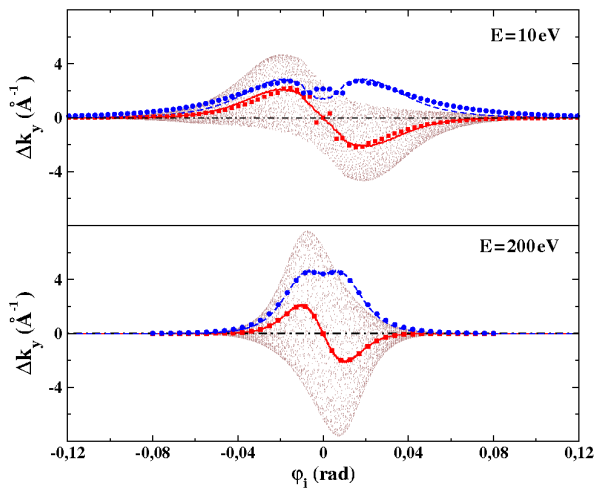}
\caption{(Color online)
$\Delta k_x=(p_{xf}-p_{xi})/\hbar$ (black dashed-dotted line) and $\Delta k_y$ (brown dots) 
versus the initial azimuthal angle $\varphi_i$, for classical trajectories closely
aligned to the $[110]$ direction $(\varphi_i=0)$. All the trajectories have incident polar 
angle $\theta_i=0.506\pi$, initial $z_i=60$\AA$\,$ and $(x_i,y_i)$ chosen at random in a unit 
cell. The red solid line and the blue dashed line represent the mean values and the 
root-mean-square values of $\Delta k_y$ respectively, obtained at each angle $\varphi_i$ from 
an average over $2000$ classical trajectories with identical initial conditions, except 
for the coordinates $(x_i,y_i)$. The mean values (red squares) and the root-mean-square values 
(blue circles) derived from the quantum diffraction probabilities are also depicted.}
\label{fig:figure1}
\end{figure}
%


Figure \ref{fig:figure1} shows the changes in $p_x$ and $p_y$ for classical trajectories closely 
aligned to an index direction. Two different behaviors are clearly distinguished; there are significant 
changes in $p_y$ in the region around the symmetry direction $\varphi_i=0$, identified 
as the quasiresonance region, and a constant value of $p_y$ outside it. The adiabatic invariant
in the quasiresonance $p_x$ remains constant in the whole depicted range. The change 
in $p_z$ (not shown in the figure) is dictated by the energy conservation in the global 
elastic process.

For the quasiresonance conditions (a) and (b), the quantum mechanical treatment 
of the collision can be simplified, as only a small number of quantum channels are
effectively open \cite{Rousseau07}. 
Under these conditions the large momentum component along the index direction remains
constant and only quantum transitions corresponding to reciprocal lattice vectors 
perpendicular to this direction present a non-negligible diffraction probability. 
As only a small amount of energy associated with the perpendicular degrees of
freedom is available, only a few quantum channels contribute to the diffraction 
spectrum. 
This effect has been experimentally confirmed for incidence along a low crystal index 
\cite{Rousseau07,Schuller07,Schuller08}.
Then the quantum dynamics can be described by an effective 2D model 
\cite{Rousseau07} with the averaged potential Eq. (\ref{vav}).

%
\begin{figure}[tb]\centering
\includegraphics[width=0.96\linewidth]{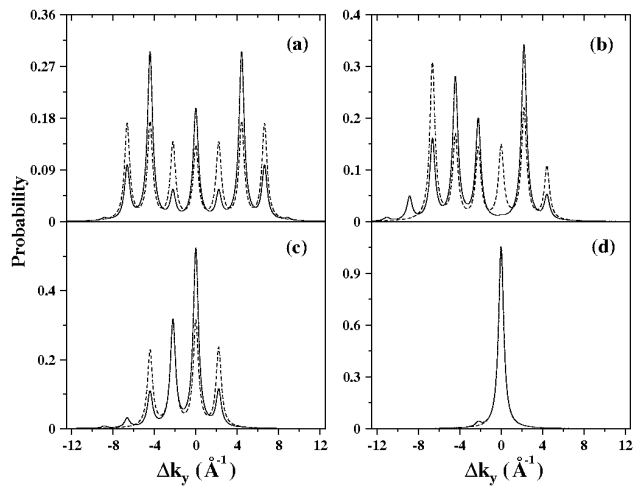}
\caption{
Quantum (solid lines) and quasiclassical (dashed lines) diffraction spectra for a grazing 
incident beam with energy $E=200\,eV$, polar angle $\theta_i=0.506\pi$ and different 
azimuthal angles $\varphi_i$; (a) along the $[110]$ direction, $\varphi_{i}=0$, (b) 
inside the quasiresonance region, $\varphi_{i}=0.01$, (c) on the edge of the quasiresonance 
region, $\varphi_{i}=0.02$, and (d) outside of the quasiresonance region, $\varphi_{i}=0.04$. 
The quasiclassical probabilities were calculated replacing the continuous distribution by an histogram.
The diffraction probabilities have been convoluted with a Lorentzian function to account for 
possible broadening effects.}
\label{fig:figure2}
\end{figure}
%

%
Figure \ref{fig:figure2} shows the diffraction spectra for four incident azimuthal angles: along a 
crystal index direction, inside, on the edge, and outside the classical quasiresonance region.  
The significant non-specular diffraction peaks  in the first three spectra correspond with the 
efficient energy transfer between the perpendicular momentum components, $p_{y}$ and $p_{z}$, 
predicted in the classical quasiresonance.
As the incident direction approaches the edge the intensity of the inelastic
channels decreases, being the specular diffraction peak the only that survives for incidence 
outside the quasiresonance region.  
The precise distribution of the diffraction probabilities is dictated by quantum effects in the 
dynamics of the system, and as expected, is not correctly reproduced by the classical model, see 
figure \ref{fig:figure2}.
The important result however is the qualitative agreement, and especially the correct prediction of a 
significant qualitative change in the quantum diffraction spectrum as the incident direction crosses 
the border of quasiresonance. Both the exact classical and exact quantum results confirm the simple 
quasiresonance analysis.   
As figure \ref{fig:figure1} shows, there is also a very good agreement between the 
trends in classical and quantum changes in $p_{y}$. 
Hence both classical and quantum dynamics reflect the same mechanism of efficient energy transfer that 
characterizes quasiresonance around the low index direction. 
  

%
\begin{figure}[tb]\centering
\includegraphics[width=0.9\linewidth]{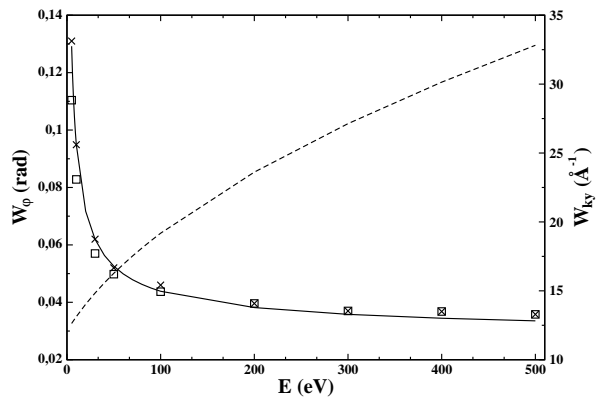}
\caption{
The span of the quasiresonance region around the $[110]$ direction versus the 
incident energy $E$, in a process with grazing incident angle $\theta_i=0.506\pi$.
Labels on the left correspond the angular width $W_\varphi$; the squares are 
the numerical classical estimation, the crosses the numerical quantum estimation,
and the solid line the classical prediction derived from (\ref{wpy}).
Labels on the right give the size of the quasiresonance region in momentum space, 
$W_{ky}=W_{py}/\hbar$; the dashed line corresponds to the classical estimation 
(\ref{wpy}). As a criterion for the numerical estimation of the angular width of the 
quasiresonance region, we consider the full width at half height of root-mean-square 
values of $\Delta k_y$.}
\label{fig:figure3}
\end{figure}
%


Figure \ref{fig:figure3} shows the dependence of the span of the quasiresonance region
on the total translational energy. The monotonic increase of the width $W_{p_y}$
with the energy is a peculiar feature of the simple interaction potential considered.
Since in this model the averaged potential is zero except for the principal surface 
directions $(1,0)$ and $(0,1)$, quasiresonance around other symmetry directions is a higher 
order perturbative effect. Therefore, the main mechanism for the eventual demise of the 
main quasiresonance regions, i.e., the overlapping between neighbouring resonance zones 
in phase space, is nearly absent.
The angular with $W_\varphi$ decreases rapidly at low energies and changes very slowly
at higher energies.
At low energies, no normal quantum inelastic channels are energetically allowed 
and, as expected, there is not good agreement between the quantum and classical dynamics.
Both results become nearly identical as the energy increases and more diffraction
channels are open.


In conclusion, the grazing angle collision of an atom with a periodic
surface for incidence  closely aligned to a (low) index direction has been 
investigated using both a classical and a quantum description.
For a large interval of incident energies the results can be analyzed in
terms of quasiresonance. This interval is limited below by low energies for
which the dynamics is not perturbative, and above by large energies with
large normal to the surface component of the momentum for which the 
atom-surface interaction is not slowly variant.
Quasiresonance is characterized in this system by the adiabatic invariance of
the momentum component along a symmetry direction of the target surface, together with 
an efficient energy transfer between the normal degrees of freedom. 
Even when the classical results do not reproduce the precise quantum diffraction 
probabilities, the averaged results are in good agreement in both dynamics,
specially for large energies.
The width of the quasiresonance region can be classically estimated from the 
evolution of the phase-space of the perturbed system in the transient process. 
The angular width $W_\varphi$ decreases rapidly at low energies and remains nearly 
constant for large energies, provided that the interaction does not activate high 
order quasiresonances.

We have also analyzed the analogies between the classical and quantum quasiresonance
behavior. The mayor disagreement between both dynamics occurs at low energies, due to
the energy threshold of the quantum diffraction channels. As was described in 
ultracold atom-diatom collisions \cite{Forrey99}, no quantitative similarity is found
when the inelastic channels are closed.
For large translational energies, the quasiresonance width coincides, indicating a 
common mechanism in both dynamics.

Our results for energies up to hundreds of eV indicate an angular quasiresonance
width $W_\varphi\approx 0.02$ rad, which is above the beam divergence achieved in
experimental setups.
Although we have assumed a perfect rigid surface, this is a good approximation for large
beam energies as the particle wavelength is much smaller that the thermal vibration
of the lattice atoms \cite{Rousseau07}.
Therefore, experimental investigations to characterize quasiresonance regions 
in more realistic atom-surface interactions, measuring the diffraction pattern 
for incident beams closely aligned to the low crystal index direction are in 
principle possible.


A. R. and J. P. P. acknowledge financial support from Spanish MCT 
(Grants No. FIS2004-05678, No. FIS2005-02886, and No. FIS2007-64018) 
and Gobierno de Canarias (Grant No. PI2004/025).


\end{document}